\documentclass[aps,prl,preprint,groupedaddress]{revtex4}
\usepackage{graphicx}
\newcommand{\be}{\begin{equation}}
\newcommand{\ee}{\end{equation}}

\begin{document}

\title{Intriguing examples of inhomogeneous broadening}
\author{Francesco Ancilotto$^{1}$, M. Mercedes Calbi$^2$, Milton W. Cole$^{2,*}$,
 Silvina M. Gatica$^{2,3}$ and E. Susana Hern\'andez$^{3}$ }

\affiliation
{
$^1$ INFM (UdR Padova and DEMOCRITOS National Simulation Center, Trieste, Italy) \\ and Dipartimento di Fisica ``G. Galilei'',
Universit\`a di Padova,
\\ 
via Marzolo 8, I-35131 Padova, Italy;\\ $^2$Physics Department, Pennsylvania State University, University Park, Pennsylvania 16802; 
\\
 $^3$ 
  Departamento de F\'{\i}sica, Facultad de Ciencias Exactas y Naturales, Universidad de Buenos Aires, and   Consejo Nacional de 
Investigaciones Cient\'{\i}ficas y T\'ecnicas, Buenos Aires, Argentina.  
\\
$^*$Corresponding author: 104 Davey Laboratory, Penn State University, University Park, PA 16802, USA; mwc@psu.edu
}


\begin{abstract}

Three problems are considered in which inhomogeneous broadening can yield unusual consequences. 
One problem involves the energy levels of atoms moving within nanopores of nearly cylindrical cross section.
A second involves atomic or molecular motion in a quasi-one dimensional interstitial channel within a bundle of carbon nanotubes.
The third  problem involves motion within a groove between two nanotubes at the surface of such a bundle.
In each case, the density of states at low energy is qualitatively different from that occurring in the perfectly homogeneous case.

\end{abstract}
\maketitle

\section{1. Introduction}

Problems involving inhomogeneous broadening are ubiquitous in physics and chemistry. The usual consequence of particles experiencing a variety of environments (or a range of velocities) is that narrow spectral lines become smeared into broader line shapes. The broadening is a nuisance in some circumstances while in others it provides valued information about the distribution of local environments. Interestingly, in some cases quite different behavior is predicted to occur. These examples pertain to the quantum states of atoms or molecules moving in somewhat heterogeneous environments. Presumably, similar behavior may occur for electrons in corresponding situations, e.g. electrons in liquids \cite{jortner}

The first example involves atoms or molecules moving in nanoporous media. Nowadays, such pores can be nearly perfect and either quasi-one dimensional (1D) and independent or interconnected. We focus on the 1D case here. The other two examples described in this paper involve bundles of carbon nanotubes; these are sometimes called ropes. Such bundles consist typically of 100 to 200 tubes, each of which is nearly straight and of diameter around 1.5 nm. To see interesting size effects in these problems, the energy levels of motion perpendicular to the $z$ axis (the tube or pore axis) should be relatively widely separated. That means that the zero-point energy (ZPE), which is of order $\hbar^2/(m a^2)$, should be large compared to the other energy scales in the problem, such as $k_B T$. Here $m$ is the mass and $a$ is the transverse dimension, both of which should be small  to reveal the size effects of interest. For example, if we discuss H$_2$ in a space with $a$= 1 nm, the temperature equivalent of the ZPE is $\hbar^2/(m a^2 k_B) \approx 1 K$.

Section 2 of this paper discusses the problem of particle motion within a nearly cylindrical pore. Under appropriate conditions, unusual behavior is found to occur for the adsorbed particles' density of states, with implications for the thermodynamic properties of the adsorbed phase. Section 3 discusses a more complicated case of hydrogen molecules moving in the interstitial channels between a triad of nanotubes. Much research has explored these states in connection with applications to hydrogen storage and isotope separation, as well as intriguing possibilities about 1D phases of matter [2-6]. Here, we describe a situation in which the heterogeneity associated with an experimental distribution of tube radii yields a novel distribution of energy levels, with implications for a predicted bose-einstein condensation (BEC) 
\cite{ourprl}. In Section 4, we address briefly a similar problem, the states of molecules strongly bound in the groove region between two nanotubes at the external surface of a bundle of nanotubes. Section 5 summarizes our results.

The evaluation of these examples will, in some cases, employ heuristic models that simplify the mathematics and clarify the results without sacrificing accuracy of the qualitative conclusions. Surprisingly often in science we are lucky and such models turn out to be more reliable than they are expected to be.

\section{2. Particles within nearly cylindrical pores}

The pores considered here are assumed to be uniform in the axial ($z$) direction, so that the energy levels of particles moving within them may be written as a sum of a kinetic energy of $z$ motion $E_z= p_z^2/(2m)$ of the particles and a transverse energy, $E_t$:

\be
E= E_z + E_t 
\ee

We first address the transverse problem by finding the energy $E_{cyl}(a)$ of a particle inside a perfectly cylindrical pore (radius $a$) and then evaluate the effects of eccentricity on the spectrum. An atom or molecule confined within a small cylindrical pore experiences a potential energy $V(r)$ as a function of radial distance $r$ that can be extremely attractive compared to that on a planar surface made of the same material. This expectation is borne out in a commonly employed simple model in which the potential is derived by integrating Lennard-Jones interactions between the particle and the individual atoms of the host medium ($r>a$). If the parameters characterizing this interaction are $\epsilon$ and $\sigma$, the resulting potential energy on the axis of the pore is

\be
U=V(0)= U_0 F(R)
\ee
with 
\be
F(R)= \frac{7}{32} \frac{1}{R^{9}}- \frac{1}{R^{3}}
\ee
and
\be
U_0 = n \pi^2 \epsilon\; \sigma^3
\ee

Here, $n$ is the number density of the host solid and $R=a/\sigma$ is the reduced radius of the pore. We are especially concerned with low-lying states in this paper. The reduced potential energy function $F(R)$ has its minimum value at $R=R_{min}=(21/32)^{1/6}\approx 0.932$, in which case $F(R_{min})\approx 0.823$.  Fig. 1 depicts $F(R)$, which exhibits this minimum, the focus of the present discussion. The energy scale for this problem is $U_0$; if we assume parameters corresponding to the H$_2$/graphite interaction ($n=0.11/$\AA$^3$, $\sigma=2.97 $\AA \ and $\epsilon =43 K$) \cite{wang1980}, then $U_0=1220 K$. This value is about four times as large as the binding energy of H$_2$ on a graphite surface and somewhat larger than the low coverage heat of adsorption measured on carbon nanotubes. \cite{vilches,footnote2}

We proceed to calculate the total energy by solving the Schrodinger equation. In the present study, for simplicity, we assume that the potential energy of the particle has a constant value $V(r)=U$ for all $r\le a$, but $V(r)=\infty$ for $r> a$. The ground state eigenvalue for this textbook problem is then given by the expression

\be
E_{cyl}(a) = U + \frac{(\hbar \alpha)^2}{2ma^2}
\ee

Here $\alpha =2.405$  is the first zero of the Bessel function of order zero (i.e., the ground state wave function). The importance of the second (kinetic energy) term for states near $R=1$ depends on a dimensionless quantum parameter $\eta= h^2  /(nm \epsilon\sigma^5)$. For the case of the present parameter set, its value is $\eta=0.02$, implying that the kinetic energy term is about 2\% of the potential energy, if $R\approx 1$. We note that this small value is a consequence of the relatively large value of $a \approx\sigma$. The resulting spread in the wave function is a factor about 10  larger than that of H$_2$ in an interstitial channel in a nanotubes bundle, discussed in Section 3 below, for which the kinetic energy is a factor $\sim 100$ greater. 

The total energy function $E_{cyl}(a)$ has its minimum value for a pore radius near $a_{min}= R_{min}$, i.e. essentially that found in the absence of the kinetic energy term. We call the corresponding minimum energy $E_{min} = E_{cyl}(a_{min})$. We will use an expansion near this minimum,

\be
E_{cyl}(a) = E_{min} + \frac{1}{2}\mu \;(1- \frac{a}{a_{min}})^2
\ee
Here, $\mu = \sigma^2 (d^2 E_{cyl}(a) /da^2)$ with the derivative evaluated at $a=a_{min}$.
Neglecting the tiny corrections due to the kinetic energy term, we find $\mu=25.6\; U_0$. 

We now consider the effect of deviation from a perfect cylinder. As an example, we treat the case of a quadrupolar distortion, for which the distance $ R$ of the pore boundary from the $z$ axis satisfies

\be
R(\phi)= a [1 + \lambda cos(2\phi) ]
\ee
(A dipolar deformation would give no energy shift, to this order). Here $\phi$ is the azimuthal angle and $\lambda$ is a parameter, assumed small, characterizing the amplitude of the distortion. Boundary condition perturbation theory \cite{fetter} yields the energy spectrum:

\be
E_t =E(a, \lambda) = E_{cyl}(a) + \frac{1}{2}\nu \lambda^2
\ee

Here, $\nu= (\hbar\alpha)^2/(ma_{min}^2) [ 1 + \alpha (d \;\ln J_2(x)/dx)_{x=\alpha} ]$, where $J_2$ is the Bessel function of order 2.
The numerical value of the quantity in brackets is $1.9$. Note that $\nu/\mu$ is of order $\eta\ll 1$; that is, the energy shift due to the deformation is comparable to the undeformed kinetic energy, a small quantity.

We now consider a particular experimental sample in which the pores' distribution is given by a specified function $N(a,\lambda)$. The transverse density of states $g(E)$ for this system is derived by integrating over these variables and finding which states have the specified energy:

\be
g(E) = \int da\; d\lambda\; N(a,\lambda) \delta[E - E(a, \lambda)]
\ee
We focus here on the lowest-lying states of the problem. These arise from those pores having radius near $a_{min}$ and small values of $\lambda$. In such a case, we may extract the corresponding distribution function as a prefactor:

\be
g(E)= N(a_{min},0) \int da \; d\lambda\; \delta[E - E(a, \lambda)]
\ee
The subsequent mathematics is straightforward if one considers very low $E$, in which case one may employ the expansions defined above to perform the integrals analytically:

\be
g(E)=\frac{\pi}{2} N(a_{min},0)\; \frac{a_{min}}{\sqrt{\mu\nu}}  \;  \Theta(E- E_{min})
\ee
Here $\Theta(x)$ is the Heaviside unit step function. This equation expresses the fact that the transverse density of states is a constant above threshold, at least over the range in which the quadratic expansions are valid. This finding may be surprising, at first sight, because one might not have expected a discrete line spectrum to be broadened into a constant, smooth continuum. A phase space argument helps to rationalize the result. The states of a given energy $E$ above $E_{min}$ correspond to the equation 
$\mu\lambda^2+\nu\delta^2=k (E- E_{min})$, where $k$ is a constant and $\delta= 1- a/a_{min}$. This is the equation of an ellipse in the  $\lambda-\delta$ parameter space. The density of states is proportional to the area between two ellipses of energy differing by $dE$, divided by $dE$. This analysis leads to the constant density of states specified above. We note that this behavior occurs for the same reason that the semiclassical quantum theory for the harmonic oscillator yields the correct constant spacing of energy levels, with its two degrees of freedom: quantization of the action integral (essentially the area enclosed by the ellipse) means that successive states differ by the constant classical frequency.

We note that the constant (i.e. energy independent) density of states for this problem coincides with that of a 2D ideal gas. It is not surprising, therefore, that when this transverse spectrum is convoluted with the longitudinal energy contribution $E_z$  
 the resulting power law behavior of 
 $N(E)$ is that of a 3D gas: $N(E)\sim (E-E_{min})^{1/2}$. Thus, hydrogen in this confined geometry exhibits low energy behavior which it would have in the absence of any confinement at all! This conclusion differs from the result for a monodisperse situation, in which case the confined particles are characterized by 1D physics, i.e. $N(E)\sim (E-E_{min})^{-1/2}$. In a more complete paper, yet to be written, we will present the implications of this interesting result for the thermodynamic properties of the system. There, we will also discuss the case of more general deformations of the form $cos(p\phi)$, where $p$ is an integer. It will be seen that the quadratic dependence of the energy on the square of the amplitude of the deformation is a general property. The effect on the density of states in that general case will differ from that described above.

\section{3. Particles within interstitial channels }

Bundles of nanotubes and individual tubes are only beginning to be characterized \cite{Weisman} and the assignments of chirality (the two integers specifying the kind of tube) and radius are usually uncertain. At present, most experimental samples have fairly broad distributions of sizes. Nevertheless, most research concerning adsorption has assumed monodisperse environments, for simplicity. An exception to that rule is a recent paper \cite{shi and Johnson}  in which heterogeneity was shown to be essential for understanding adsorption isotherm data \cite{experiments} obtained for various gases within bundles. Here, we consider the effects of such heterogeneity on the energy spectrum of H$_2$ molecules confined to the interstitial channel (IC) region between the tubes. The key function of interest in this analysis is the density of states $N(E)$. As in the previous discussion, the assumption of translational invariance in the $z$ direction (accurate for nanotubes) means that the total energy of the molecule may be written as a sum of a transverse energy ($E_t$) and a longitudinal energy $E_z  =p^2/(2m)$, where $p$ is the 1D variable momentum. 

The interesting problem is the transverse energy, which is very sensitive to the radii ($R_1$, $R_2$ and $R_3$) of the neighboring tubes. We skip the details of this calculation, which parallels that of the previous section; a complete description will be published elsewhere. The key points to understand are the following. First, there exists a particular combination of nanotubes for which the total molecular energy is a global minimum, $E_{min}$. This occurs for tubes of equal radius 
$\{{R_i}\}=R_{min}=9.95 $ \AA, in which case $E_{min}= -1053 K$. As in the previous section, we first evaluate the transverse density of states $g(E)$ of the
molecule. At low energy, the relevant states arise from ICs with radii values
close to $R_{min}$. To evaluate $g(E)$, we determine the iso-energy contours in $\bf R$
space, in which a particular IC is represented by a point with coordinates
equal to the three radii. Figure 2 depicts the dependence of the molecule's
energy on a planar surface in $\bf R$ space that passes through the point $ R_{min}(1,1,1)$. One observes that these contours are extremely anisotropic; the reason is that the energy difference between the global minimum energy and that of the point ${\bf R}=R_{min}(1,1+x,1-x)$ is a quadratic function of $x$ with an extremely small coefficient. This gives rise to the very narrow valleys observed in Figure 2.

In computing $g(E)$, one proceeds to evaluate the number of ICs between energy contours at $E$ and $E+dE$. We find the result that $g(E)$ is proportional to $(E - E_{min})^{1/2}$; the power 1/2 occurs because $\bf R$ space is three-dimensional. (The general behavior in d dimensions is a power $(d/2-1)$; the previous section exemplified $d=2$). Note that $g(E)$ is mathematically coincident with the density of states of a 3D gas. This means that inclusion of motion in the $z$ direction gives rise to behavior consistent with an ideal gas in 4D: $N(E)$ is proportional to $(E-E_{min})$. Such a spectrum yields an extraordinary consequence: parahydrogen, the low temperature (T) form of hydrogen, is therefore predicted to undergo a bose-einstein condensation (BEC) when confined to this environment, a bundle of nanotubes differing in size. The mathematics of this analysis follows that of the textbook derivation of BEC in 3D.\cite{pathria} Figure 3 depicts the resulting transition temperature as a function of the molecular number density. The typical values are of order milli-Kelvin, easily achievable in a low temperature laboratory experiment. 

This predicted transition is remarkable because it is a consequence of the system's heterogeneity. If all of the ICs were identical, in contrast, the description would be that of a 1D gas, which exhibits no transition. While unusual, this existence of a transition only in the inhomogeneous case is not unique. There exist other phase transitions that occur as a consequence of nonuniformity, an example of which is the spin glass transition.

We note several implicit assumptions in the present analysis. One is the neglect of intermolecular interactions; this assumption may be valid because of significant screening of the intermolecular interactions by the nanotubes \cite{screening}. An expected singular heat capacity at the transition (described elsewhere) relies on the assumption that molecules within an array of ICs can redistribute themselves as a function of T. This equilibration may be difficult to observe due to the low mobility of the molecules within the ICs. An additional assumption is the existence of a continuous distribution of radius values. The radius values of nanotubes are determined by the so-called chiral integers (n,m) characterizing the tubes. The spacing between consecutive values of radius is less than 0.002 nm, so the discreteness should not be a problem. \cite{footnote} Evidently, the predicted observation of BEC for hydrogen molecules, not seen in any experiment thus far, would be exciting! \cite{footnote3}

\section{ 4. Particles in the groove}

The third geometry of interest here is that of the groove region between two nanotubes, e.g. at the external surface of a bundle of tubes. We have analyzed this problem by diagonalizing the anisotropic dynamical equations of molecular motion in the vicinity of the equilibrium position to obtain the two harmonic frequencies ($\omega_1$ and $\omega_2$) characterizing this motion, as a function of the two radii. 

Figures 4 and 5 show how these frequencies vary as a function of the radius values. Note that the average  frequency exhibits a maximum near the point $R_2 = R_1  = 2$\AA. This means that the study of the vibrational spectra of these molecules would exhibit behavior characteristic of a maximum in the vicinity of the frequency corresponding to a double excitation (i.e., at $\omega_{sum}= \omega_1 + \omega_2 $). Without going into detail, this situation is mathematically equivalent to that described in Section 2, since that description involves two parameters. Thus, the line shape for an absorption experiment would be constant just below the threshold at $\omega_{sum}$, above which no absorption occurs (at least from this combination mode). This is to be contrasted with the narrow line present in the absence of heterogeneity.

\section{ 5. Summary}

In this paper, we have described a number of inhomogeneous systems that exhibit interesting physical properties. The common theme discussed here is that the nonuniformity leads to anomalous behavior that is qualitatively different from that of the uniform system. In some respects, the behavior is much more interesting than that of the uniform system. For the physical systems explored here, such heterogeneity is the rule rather than the exception. Hence, one should resist the temptation to ignore heterogeneity.

 Joshua Jortner has been an inspiration to one of us (MWC) for many years. We are grateful to the National Foundation for support of this research;
F.A. acknowledges funding from MIUR-COFIN 2001; E.S.H. acknowledges hospitality at the Department of Physics of PSU where this work was done.

\newpage

\begin{figure*}
\includegraphics[height=4in]{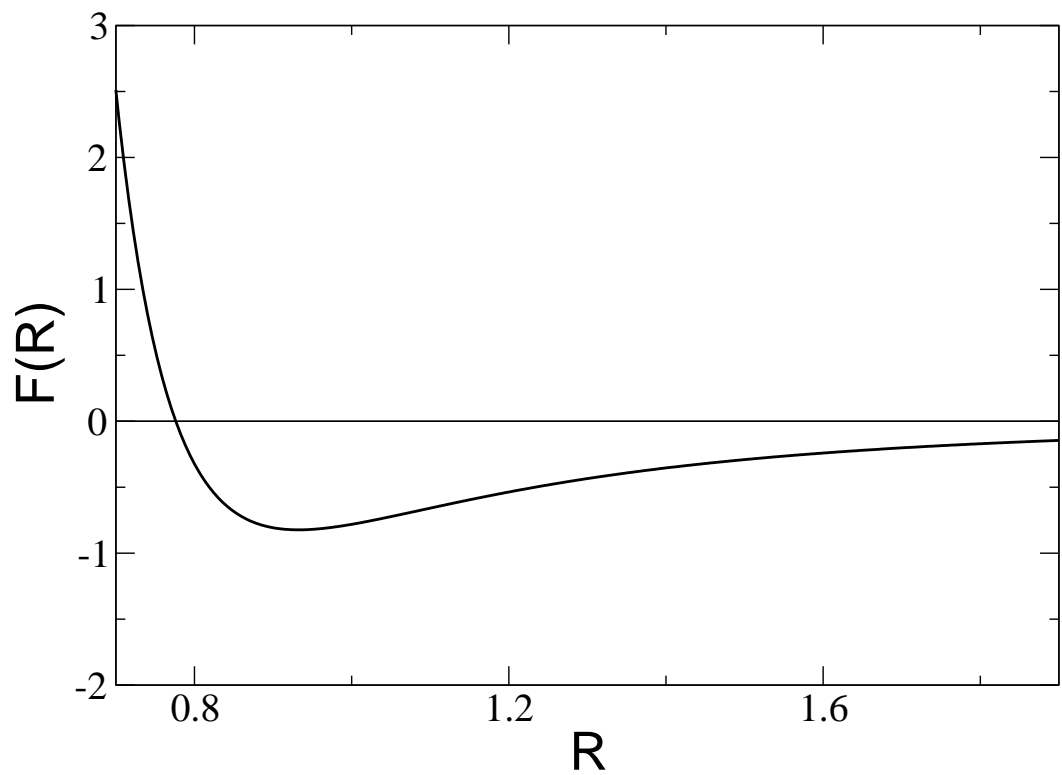}
\caption{ Reduced potential energy of a molecule in a pore, defined in Eqs. 2 and 3, as a function of the reduced radius of the pore $R=a/\sigma$. }
\end{figure*}

\begin{figure*}
\includegraphics[height=4in]{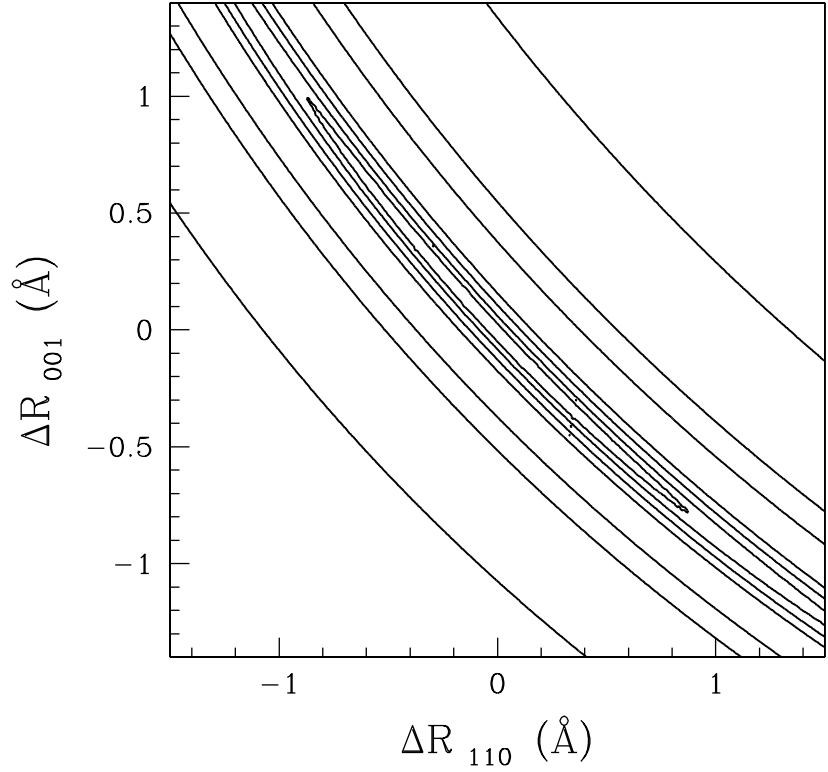}
\caption{  Contour plot showing the variation $E_t({\bf R})-E_{min}$ near the 
minimum ${\bf R}_{min}$ (which is chosen as the origin). From the contour closest to the 
minimum (the closed one) to the more distant ones, the energy contours correspond to 
$E_t({\bf R})-E_{min} = 0.005, 0.01, 0.05, 0.1, 0.5, 1$ and 5 K. }
\end{figure*}

\begin{figure*}
\includegraphics[height=4in]{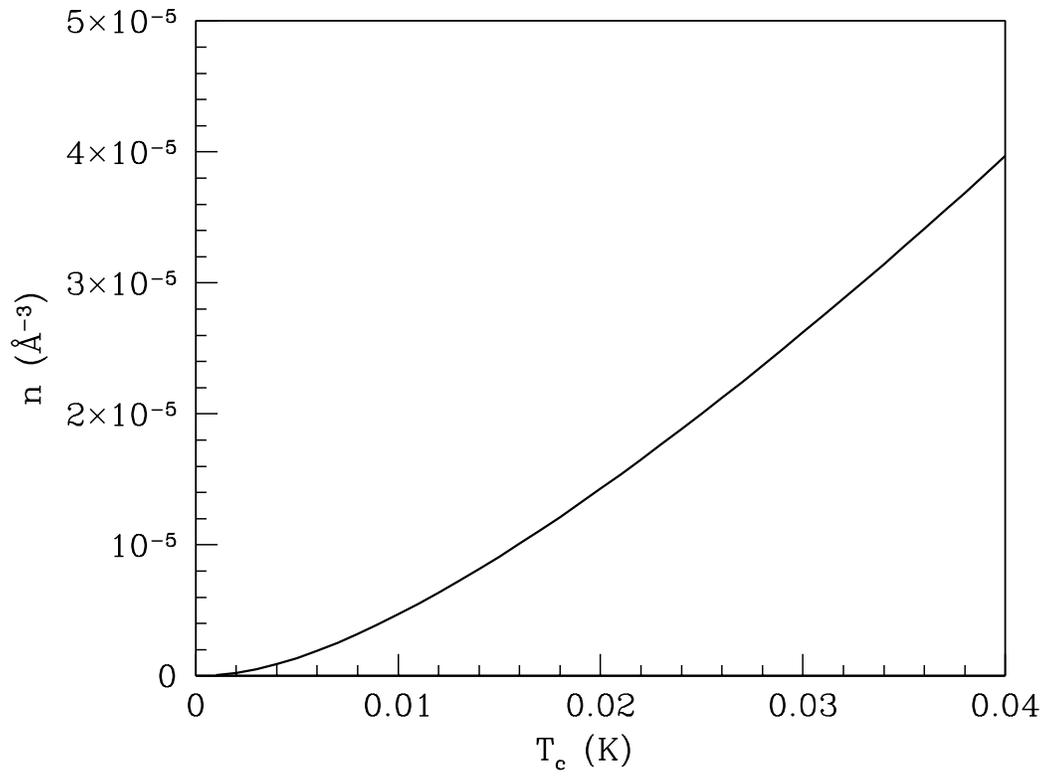}
\caption{Density of H$_2$ molecules as a function of BEC transition temperature.  }
\end{figure*}

\begin{figure*}
\includegraphics[height=5in,angle=-90]{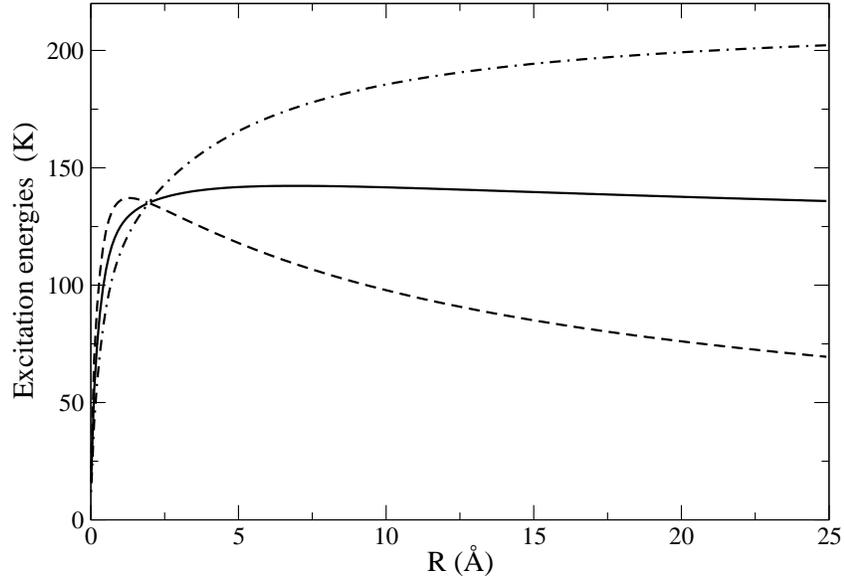}
\caption{ Excitation energies (in K) $\hbar \omega_1$, 
$\hbar \omega_2$  of one
 H$_2$ molecule in the potential well of the groove (dashed and dot-dashed 
lines, respectively) together with the ground state eigenvalue 
$\hbar (\omega_1 + \omega_2)/2$, full curve, as functions of radius $R$ (in \AA) for
the symmetric configuration $R = R'$. }
\end{figure*}

\begin{figure*}
\includegraphics[height=7in]{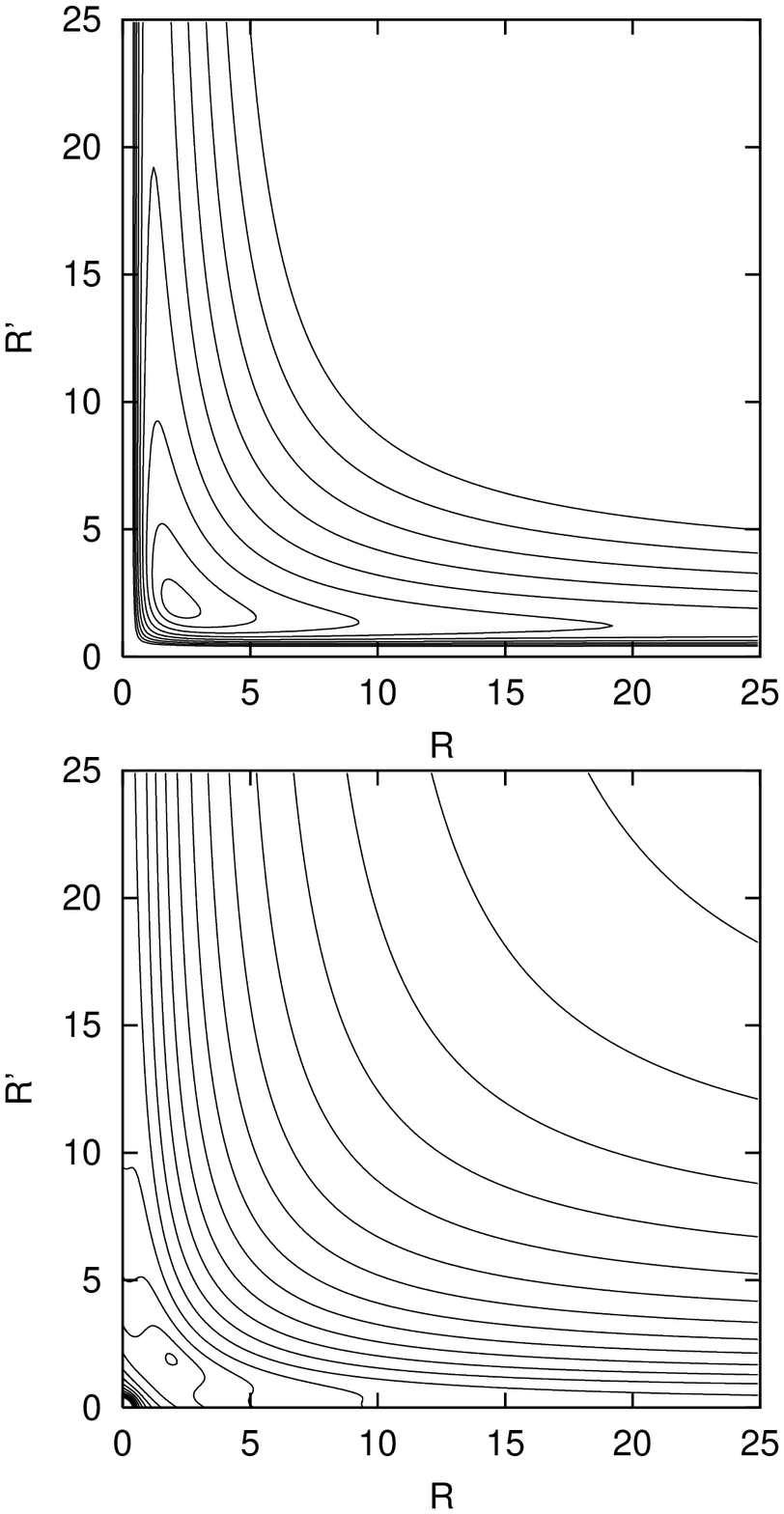}
\caption{ 
Isoenergy contours of the excitation energies 
$\hbar \omega_1$, $\hbar \omega_2$ (upper and lower panels, respectively)  as 
functions of radii $R, R'$ (in \AA). Contours  run between 100 K and 200 K in 
steps of 4K in both graphs.}
\end{figure*}


\begin{references}

\bibitem{jortner} B.E. Springett, M. H. Cohen and J. Jortner,
J. Chem. Phys. 48, 2720 (1968).

\bibitem{migone} S. Talapatra, D.S. Rawat and A.D. Migone, 
J.of Nanoscience and  Nanotechnology
{\bf 2}, 467 (2002).

\bibitem{rmp}    M.M. Calbi, M.W. Cole, S.M. Gatica, M.J. Bojan and G. Stan, Rev. Mod. Phys {\bf 73}, 857 (2001). 

\bibitem{vilches} T. Wilson, A. Tyburski,M.R. DePies, O.E. Vilches, D. Becquet, M. Bienfait 
J. Low Temp.  Phys, {\bf 126}, 403 (2002).



\bibitem{hallock} Y.H. Kahng, R.B. Hallock and E. Dujardin 
Phys. B {\bf 329}, 280 (2003).

\bibitem{wei}  B.Y.Wei, M.C. Hsu, Y.S. Yang, S.H. Chien and H.M. Lin, 
Mat.  Chem. Phys.
{\bf 81}, 126 (2003).

\bibitem{ourprl} F. Ancilotto, M.M. Calbi, S.M. Gatica and M.W. Cole, submitted to Phys. Rev. Lett.

\bibitem{wang1980} C. Wang and R. Gomer, Surf. Sci. {\bf 91}, 533 (1980).

\bibitem{footnote2} A calculation of the adsorption potential on a flat surface, using the same model, yields a potential energy of well depth $D=\frac{4\pi}{9}(\frac{5}{2})^{1/2}n\epsilon\sigma^3$. This is smaller by a factor 3.7 than the cylindrical pore minimum. 

\bibitem{fetter} Alexander L. Fetter, John Dirk Walecka, {\it Theoretical mechanics of particles and continua.}    McGraw-Hill, New York (1980).

\bibitem{Weisman} S.M. Bachilo, M.S, Strano, C. Kittrell, R.H. Hauge, R.E. Smalley and R. B. Weisman  , Science {\bf 298}, 2361 (2002).

\bibitem{shi and Johnson} W. Shi and J. K. Johnson,   Phys. Rev. Lett. {\bf 91}, 015504 (2003). 


\bibitem{experiments} S. Talapatra and A.D. Migone, Phys. Rev. B {\bf 65}, 045416 (2002); 
A.J. Zambano, S. Talapatra and A.D. Migone, Phys. Rev. B {\bf 64}, 075415 (2001).

\bibitem{pathria} R.K. Pathria, {\it Statitical Mechanics.}, Butterworth-Heinemann, Oxford (1996).



\bibitem{screening}M.K.Kostov, J.C.Lewis and M.W.Cole, in Condensed Matter Theories, 
Vol. 16, edited by S. Hern\'andez  and J. Clark, Nova Science Publishers, NY, 2001, 
pp.161, http://xxx.lanl.gov/abs/cond-mat/0010015.

\bibitem{footnote} Moreover, tubes near the center of a bundle are compressed relative to tubes near the outside, providing additional variation of radius values.

\bibitem{footnote3} Other predictions of and searches for BEC of H$_2$ have been made. See for instance: 
        S. Grebenev, B. Sartakov, J. P. Toennies and A. F. Vilesov,
Science, {\bf 289}, 1532 (2000);
%
V. Ginzburg and A. Sobyanin, 
JETP Lett. {\bf 15}, 242 (1972);
%
M.C.Gordillo and D. M. Ceperley, 
Phys. Rev.
Lett. {\bf 79}, 3010 (1997);
%
O.E.Vilches, J. Low Temp. Phys. 89, 267 (1992);
%
H. J. Maris, G. M. Seidel, T. E. Huber, J. Low Temp. Phys. {\bf 51}, 471
(1983);
%
M.C.Gordillo and D. M. Ceperley, 
Phys. Rev. B {\bf 65}, 174527 (2002).





\end{references}
\end{document}